# Near-on-Demand Mobility. The Benefits of User Flexibility for Ride-Pooling Services

Zhenliang Ma, Haris N. Koutsopoulos

*Abstract*—Mobility-On-Demand (MoD) services have been transforming the urban mobility ecosystem. However, they raise a lot of concerns for their impact on congestion, Vehicle Miles Traveled (VMT), and competition with transit. There are also questions about their long-term survival because of inherent inefficiencies in their operations. Considering the popularity of the MoD services, increasing ride-pooling is an important means to address these concerns. Shareability depends not only on riders' attitudes and preferences but also on operating models deployed by providers. The paper introduces an advance requests operating model for ride pooling where users may request rides at least *H* minutes in advance of their desired departure times. A platform with efficient algorithms for request matching, vehicle routing, rebalancing, and flexible user preferences is developed. A large-scale transportation network company dataset is used to evaluate the performance of the advance requests system relative to current practices. The impacts of various design aspects of the system (e.g. advance requests horizon, vehicle capacity) on its performance are investigated. The sensitivity of the results to user preferences in terms of the level of service (time to be served and excess trip time), willingness to share and place requests in advance, and traffic conditions are explored. The results suggest that significant benefits in terms of sustainability, level of service, and fleet utilization can be realized when advance requests are along with an increased willingness to share. Furthermore, even "near-on-demand" (relative short advance planning horizons) operations can offer many benefits for all stakeholders involved (passengers, operators, and cities).

Keywords: on-demand mobility; ride pooling; advance requests; near-on-demand mobility; rebalancing; vehicle miles traveled

## I. INTRODUCTION

Innovative Mobility-on-Demand (MoD) services have been growing rapidly, including ride-hailing or Transportation Network Companies (TNC), car sharing, micro-transit (as well as bike and scooter-sharing services). They are convenient to use and provide a good level of service (LOS) to users, both, in space and time, compared to not only transit but also traditional taxi services. They also pose significant challenges to transport policymakers, planners, and operators.

While MoD services offer the promise to improve mobility, this "on-demand" convenience possibly comes at a cost as they also raise concerns regarding their impact on congestion and transit ridership. For example, the San Francisco County Transportation Authority (SFCTA) reported in 2017 that, during peak hours, TNC vehicles are 15 times more than taxi vehicles, and TNC trips are concentrated in the congested areas comprising 20-26% of all trips in these areas. The daily TNC driving miles (in-service and out-of-service) within San Francisco are estimated to account for 20% of all the local daily miles [1, 2]. At the same time, many transit systems experience a reduction in ridership in recent years, especially buses. Chicago Transit Authority (CTA), for example, reported that the decline of ridership was partly caused by TNCs competition. Equally alarming is the student ridership decline despite the free travel pass [3].

There are also concerns about the TNC companies' long-term financial survival because of inherent inefficiencies in the current operating models combined with aggressive pricing strategies. Currie and Fournier [4] compiled a database of 120 systems, including dial-a-ride (DAR), demand responsive transit (DRT), and Micro-transit, since the 1970s. They found that 50% last less than 7 years, 40% last less than 3 years, and about a quarter fail within 2 years. High operating costs are the main contributor to failure services.

Considering the popularity of the MoD services, increasing ride-pooling (sharing a vehicle ride with others) is an important means to reduce Vehicle Miles Traveled (VMT), an important objective for sustainable transportation, as it relates directly to reduced emissions, improved traffic conditions, and energy savings. Increasing ride-pooling also has the potential to not only reduce congestion and environmental footprint, as trips could be covered at reduced VMT, but also improve operating efficiency, which is of great interest to TNCs themselves (considering that so far most of these companies are operating at a deficit) [4, 5]. TNCs have indeed started to focus on ways to promote ride-pooling through services such as Uber Pool, Uber express, and Lyft Line [6]. Lyft redesigned its app and is developing strategies to improve ride pooling [7] and, most recently, is expanding the 'Wait & Save' pilot in response to COVID-19 [8]. Riders choosing 'Wait & Save' pay less than a standard Lyft ride, and the longer they wait, the more they save.

Various studies in the literature also point to the value of ride-pooling. Using taxi trip data from New York City, Santi, et al. [9] showed that even if passengers are willing to share trips with just another passenger, a significant benefit in VMT can be achieved in dense metropolitan areas, such as Manhattan. Alonso-Mora, et al. [10] showed that if all passengers are willing to share trips, 25% of existing taxis (seating capacity 4) in Manhattan can satisfy 98% of the ride requests with an average wait time of 2.8 minutes and mean trip delay of 3.5 minutes. Many studies looking at the impact of autonomous ride-pooling on traffic congestion, environmental costs, and users' travel choices [9-12], highlight the potential of ride-pooling to reduce costs (e.g. VMT) and improve service efficiency.

Studies have also explored the interactions among factors impacting ride-pooling using real-world data. Tachet, et al. [13] presented a simple index that measures the potential for ride pooling as a function of the geographical area, intensity of requests, average travel speed in the area, and LOS constraints (users tolerance such as wait time). They also show that various metropolitan areas (New York, San Francisco, Singapore, and Vienna) exhibit similar behavior in terms of shareability. The generality of the results implies that the same operating strategies to increase shared rides could work in metropolitan areas with different characteristics and structure.

In general, the potential to increase pooled trips is determined by four factors:

Z. Ma is with the Institute of Transport Studies (ITS), Department of Civil Engineering, Monash University, Melbourne, VIC 3800, Australia (e-mail: mike.ma@moansh.edu).
H. N. Koutsopoulos is with the Department of Civil and Environmental Engineering, Northeastern University, Boston, MA 02115, United States (e-mail: h.koutsopoulos@northeastern.edu).

- Operating and scheduling strategies to facilitate ride-pooling and take advantage of the available opportunities for pooling requests to the full extent (request matching, vehicle dispatching, and rebalancing, etc.).
- Users' willingness to share (demand for ride-pooling) and LOS expectations (extra ride time due to sharing, waiting for a vehicle, etc.).
- Spatiotemporal distribution of requests, which determines the opportunity to match sharing requests.
- Market fragmentation (various service providers) and competition, which impact the sharing of information and resources (leading to inefficient resource utilization).

The paper focuses on the first factor, exploring alternative operating strategies to increase pooling as a means to reduce network impacts and improve productivity. An important challenge in improving current operations is the uncertainty with future demand. Central to on-demand ride-pooling operating practices is the term 'on-demand', which characterizes the majority of the TNC services currently offered. The term implies that the service is available when it is requested. Users request the service when they need it and expect it to be available almost immediately. Hence, the time between the placing of the request and the initiation of the service is relatively short. Clearly, from a scheduling point of view, this lack of knowledge of future requests limits the ability of the scheduling and matching algorithms to take full advantage of shareability opportunities.

Approaches to deal with the lack of knowledge of future requests in order to increase pooling opportunities are twofold:
a. using information about future demand (e.g. predicted demand) when scheduling and routing decisions are being made; and
b. deploying operating models that hedge future uncertainty by adding flexibility in the operations, as algorithmic and modeling innovations alone may not be able to address the problem.

Routing decisions with some knowledge of future requests have been shown to improve MoD performance [14-18]. Using NY taxi data, Alonso-Mora, et al. [16] argue that sampling a few future requests from the distribution of historical requests can improve vehicle scheduling, reducing wait time, and trip detour time. However, it comes at the cost of a higher VMT and a lower ratio of trips actually shared (80% compared to 90% in the case without prediction). This could be caused by the fact that the vehicles are routed towards the predicted demand that may or may not be realized. Another relevant study estimated the percentage of shared trips using an oracle model with exact knowledge of current and all future taxi requests [9]. The results using NY taxi data show that with a maximum delay time of 1 minute and capacity 2 passengers per vehicle, the oracle model with infinite horizon results in 94.5% of the trips being shared compared to less than 30% shared requests in the case of purely on-demand services (1 min horizon). Using TNC data from three US cities (San Francisco, New York, and Los Angeles), Chen, et al. [14] report that ride-pooling with advanced knowledge of future requests could yield significant benefits. They mention that a system where all requests for the next 5 min are known, and vehicle capacity is 3 can pool close to 50% of all ride requests, reduce fuel consumption by 15%, and fleet size by 30%. Bertsimas, et al. [18] developed an online routing algorithm for on-demand taxi services with solo trips. They report, using the NY taxi data, that asking passengers to request service 10 minutes beforehand results in 18% increase in profit compared to on-demand requests. Tsao, et al. [19] presented a model predictive control (MPC) based ride-pooling algorithm (capacity of 2). They found, using San Francisco taxi data, that the MPC based ride-pooling system improves the LOS (wait and journey times) with a slight increase in VMT, compared to the reactive system.

Table I summarizes the findings in terms of improvements relative to the pure on-demand case when prediction (or an oracle model) of future requests is used. The % of performance changes is calculated using the finding reported in the papers [1] (it should be noted that the studies have focused on different design dimensions and reported varied performance metrics).

TABLE I
FINDINGS ON THE PERFORMANCE OF MOD SYSTEM WITH FUTURE REQUESTS

| Ref. | Objective | Future requests | Horizon | Benefit | Share | WT | DT |
|---|---|---|---|---|---|---|---|
| [9] | TT | exact | Infinite | -8.0%[a] | +11% | N/A | N/A |
| [14] | TT | exact | 5 min | -15.8%[b] | +47% | N/A | 2 min |
| [15] | TT | predicted | Infinite | -5.5%[a] | N/A | N/A | -29% |
| [16] | TT | predicted | 30 min | +12.5%[b] | -3.0% | +25% | +20% |
| [17] | WT | predicted | 2 hours | -37%[c] | N/A | -89% | N/A |
| [18] | Profit | exact | 10 min | +18%[d] | N/A | N/A | N/A |
| [19] | TT, VMT | exact | 2.5 hours | +1.3%[b] | N/A | -34% | -8.7% |

Note: N/A represents no reported measure for the study. **Objective** is the objective of the optimization, **TT** travel time, **WT** wait time, **DT** trip delay, **Share** the % of shared trips. In **Benefit**, a is total travel times, b is total VMT, c is the number of rebalancing vehicles, and d is profit.

However, algorithmic and modeling advances may not be enough and have to be combined with alternative operating models to deal effectively with future uncertainty. The second approach to deal with the impact of future demand uncertainty in order to increase shared trips aims at using more flexibility from the user side to compensate for the lack of knowledge of future requests. New operating models to increase pooling opportunities proposed in the literature include meeting points at origins and destinations [20], dynamic waiting [6], transfer points to switch vehicles [21], and integration of the on-demand services with regular transit services [22, 23]. However, such approaches have to address the fundamental challenge of increasing the actual trips being shared while maintaining a high level of service.

This paper, building on the findings from the literature (Table I), proposes an Advance Requests Ride Pooling (ARRP) model with service constraints as a means to reduce the impact of the uncertainty in future demand. "Advance requests" means that a user places a request at least $H$ minutes before the desired departure time. $H$ is referred to as the planning horizon. Users may choose "on-demand" or "in-advance" services. Service constraints guarantee that customer pick-ups and drop-offs take place within prespecified time windows. The vehicle schedule is optimized every decision epoch, e.g. every 30 seconds, in response to both the on-demand and advance requests. The price of the different services may be different based for example, on how long in advance a request may be required. Pricing plays a critical role in implementing such systems, incentivizing users to adopt the service and plan trips ahead. It is an important aspect of the problem but outside the scope of the paper.

---

[1] Data are from: [9] Figure 3, delay 5 minutes, capacity 2; [14] Table 1; [15] Figure 4, arrival rate 27 passengers/hour; [16] Figure 4, 2000 vehicles, capacity 4, 400 samples; [17] Table 1, MPC-LSTM; [18] Figure 5; [19] Table 1

Knowledge of (near) future requests, if such a service is deployed, can impact all aspects of the operations of ride pooling services, not just increasing sharing opportunities, but benefit all stakeholders involved:

- **Matching requests**. As more information is available on future requests, the routing algorithms can take advantage of this information to better organize shared rides.
- **Routing decisions**. Trip sequencing and vehicle routing also benefit from the knowledge of requests in advance, potentially leading to lower VMT.
- **Management of idle vehicles and fleet size**. Compared to pure reactive operations when only on-demand requests are available, with requests in advance, rebalancing of idle vehicles is more informed and can avoid myopic decisions (e.g. rebalance a vehicle to another location while a request is coming at the current location). Increased operating efficiency can also lead to reduced fleet size requirements.
- **LOS**. With more informed matching, routing, and rebalancing decisions, services with advance requests may also improve the level of service. While under on-demand services sharing in general increases wait time and delay, the more efficient operations when requests are placed in advance can have less impact on the passengers in terms of their wait times and possible delay.

Despite this potential, the problem has received limited attention in the literature. As Table I indicates, there is a lack of systematic studies of the performance of large-scale ride pooling services with advance requests. Hence, there is a lack of understanding of the potential performance and how various factors, such as user characteristics and preferences, impact the performance. Understanding the role of various design aspects, e.g. advance requests horizon and vehicle capacity, is also important. The closest study is the one reported in [9]. However, it considers only sharing of at most two requests, infinite planning horizon, and there is no investigation of the impact of the important factors mentioned above.

The main objective of the paper is to address these questions and shed light on the potential of a system with requests in advance as the operating mode. The empirical analysis and the findings are applicable to centrally controlled TNCs and automated MoD systems (although advanced requests can be deployed by any current service). Key contributions include:

- Introducing and exploring, through a large TNC dataset, the performance of ride pooling systems with requests in advance and how it compares to typical MoD services.
- Assessing the impact of various design aspects of the advance requests system on its performance (e.g. planning horizon, vehicle capacity, etc.), as well as user preferences (willingness to share and plan ahead), LOS constraints, and traffic conditions. The results show that "**near-on-demand**" services (relative short planning horizon $H$) can capture many of the benefits for all stakeholders (passengers, operators, and cities).
- Developing a general platform for ride pooling operations that facilitates the comprehensive evaluation of a wide range of TNC designs. The platform includes algorithms for request matching, vehicle routing, and rebalancing.

The remainder of the paper is organized as follows: Section II introduces the problem and develops a general platform for its evaluation. Section III formulates the ARRP problem and discusses algorithmic approaches for the solution of the associated assignment and rebalancing problems. An extensive case study using a large-scale TNC dataset from a city in China is presented in Section IV. Finally, the main findings and future directions are summarized in Section V.

## II. OVERVIEW AND PRELIMINARIES

The system operates under a policy of advance requests. Requests are placed ahead of time with desired pick-up time windows. In general, the service can accommodate users with different advance request horizons and charge different prices accordingly. Advance requests are expected to increase the likelihood of sharing trips and reduce operating costs. Hence, the pricing of the service can be the means of incentivizing users to make requests in advance.

Let $H$ be the advance requests horizon. Scheduling decisions are made on a rolling horizon every $\Delta t$ (decision epoch), with $\Delta t \leq H$ (Figure 1). That means that requests are scheduled at decision epochs, i.e. every $\Delta t$, utilizing all available information about on-demand and advance requests.

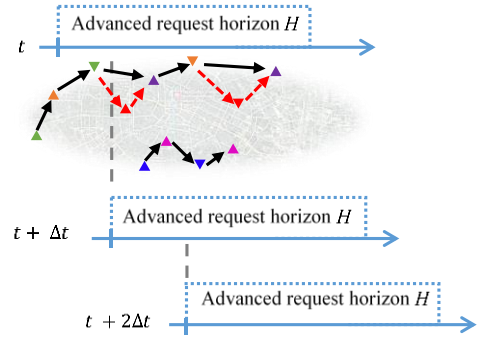

**Fig. 1.** ARRP operating model

Figure 2 illustrates the decision support framework. The decisions respect requests that have already been assigned to trips. Unfilled requests are processed by the scheduling engine and "optimally" assigned to an existing trip or form new trips or are rejected. At each decision epoch, all advance requests within the time horizon $H$ are known. The scheduling engine processes the requests respecting the characteristics of the customers in terms of their LOS preferences and service types, as well as the operating characteristics of the vehicles and prevailing traffic conditions. The scheduler makes rebalancing decisions to reposition vehicles anticipating future requests.

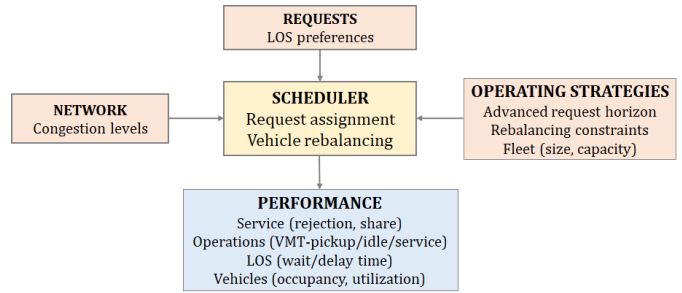

**Fig. 2.** ARRP decision support framework

The platform (Figures 1 and 2) is general and able to represent alternative operating models. For example, for $H = 0$ the model represents on-demand operating practices where users make requests expecting immediate service. In this case, the scheduling of the services is reactive with requests collected during a short time window, e.g. 30 seconds. There is a fleet of vehicles $\mathcal{M} = \{1,2,\dots,M\}$, each with finite capacity $\Omega_m$ (not necessarily the same) operating to serve a set of requests $\mathcal{N} = \{1,2,\dots,N\}$ occurring in the day. By varying parameters (e.g. planning horizon), the efficiency of alternative operating models can be assessed.

Algorithm 1 summarizes the overall approach: at each decision epoch $k$, the system is characterized by its state $x_k$ defined by the status of vehicles (location, available capacity, schedule), requests (served or not), and rebalancing zones (request rates and the number of vehicles). The decision support system pools together the outstanding requests $\mathcal{N}_k^0$ from period $k-1$ (not served yet), on-demand requests $\mathcal{N}_k^\Delta$ and advance requests $\mathcal{N}_k^H$ for the horizon $H$. Given the system state $x_k$ and zonal request rate $\Lambda$, request assignments and vehicle rebalancing decisions are made in Steps 2 and 3. Step 4 updates the system state given the assignment and rebalancing actions. The process is repeated every decision epoch., e.g. 30 seconds.

**Algorithm 1** (ARRP)

**Inputs**: $\mathcal{N}, \mathcal{M}, \Lambda$

**For** each decision epoch $k \in K$
  1: $\mathcal{N}_k \leftarrow \text{PoolRequests}(\mathcal{N}_k^0, \mathcal{N}_k^\Delta, \mathcal{N}_k^H)$
  2: $\mathcal{A}(x_k) \leftarrow \text{AssignRequests}(x_k, \mathcal{N}_k, \mathcal{M}_k)$
  3: $\mathcal{B}(x_k) \leftarrow \text{RebalanceVehicles}(x_k, \mathcal{M}_k, \Lambda)$
  4: $x_{k+1} \leftarrow \text{UpdateState}(x_k, \mathcal{A}(x_k), \mathcal{B}(x_k))$

The critical steps of Algorithm 1 are assigning requests to vehicles (Step 2) and rebalancing idle vehicles to zones (Step 3). Figure 3 illustrates an example of the current system state, request assignments, and vehicle rebalancing, with both on-demand and advance requests. Matching requests to vehicles, routing vehicles to serve requests, and rebalancing benefit from the advance requests information. For example, let us assume that the desired departure time for the requests placed in advance ($n_4$ and $n_5$) is in the next 5 minutes. If the system operates without advance requests, only requests $n_1$, $n_2$, and $n_3$ are known. Vehicle $m_1$ picks up $n_2$ and $n_1$. Vehicle $m_2$ would be assigned to serve $n_3$ and routed along the grey-dashed path (Figure 3b), vehicle $m_3$ is rebalanced to $z_2$ and $m_4$ to $z_3$ (Figure 3c). But, with advance requests, vehicle $m_2$ will serve $n_3$ and travel following the black-dashed path to serve $n_4$ (Figure 3b). Vehicle $m_3$ stays at the current location to serve the nearby advance request $n_5$ rather than rebalancing to $z_2$.

### III. METHODOLOGY

The network is represented by a set of nodes $\mathcal{I} = \{1,2,\dots,I\}$, links, and link attributes (e.g. travel time). The (shortest path) travel time and distance between any two nodes $i$ and $i'$ are denoted by $\tau_{i,i'}$ and $u_{i,i'}$, respectively, and can be the time of day specific. For rebalancing, the network is partitioned into zones $\mathcal{Z} = \{1,2,\dots,Z\}$. $\Lambda = \{\lambda_{zk} | \forall z \in \mathcal{Z}, \forall k \in K\}$ is the set of request rates. $\lambda_{zk}$ is the request rate in zone $z$ at time $t_k$.

Associated with a request $n \in \mathcal{N}$ is an OD pair $(i_n, i_{n+N})$, request time $t_n$, number of passengers $q_n$ in the request, service types $\gamma_n$ (shared or single trip, and on-demand or in advance), and LOS constraints (the desired pick-up time window $[t_n^e, t_n^l]$ and the acceptable trip delay time $D_n$) [2]. Trip delay time $D_n$ is the difference between the actual trip time and the time if the request was served as a solo trip. Customers may have different flexibility characteristics, such as service types, willingness to share and place requests in advance, etc.. For example, some customers may request a solo ride service and have strict LOS constraints, while others may request a shared ride and are flexible with respect to their waiting time window. They may place requests on-demand or in advance.

The pickup of request $n$ occurs at or after the earliest time $t_n^e$, but no later than the latest time $t_n^l$. The restriction on picking up customer $n$ at or after time $t_n^e$ means that a vehicle may need to wait at the requested pick-up location if arriving early. Let $w_{mn}$ be the wait time of vehicle $m$ (if arriving earlier) if assigned to serve request $n$. There is a maximum allowable wait time $W_m$ for $m$. Thus, a feasible assignment has to satisfy $w_{mn} \leq W_m$.

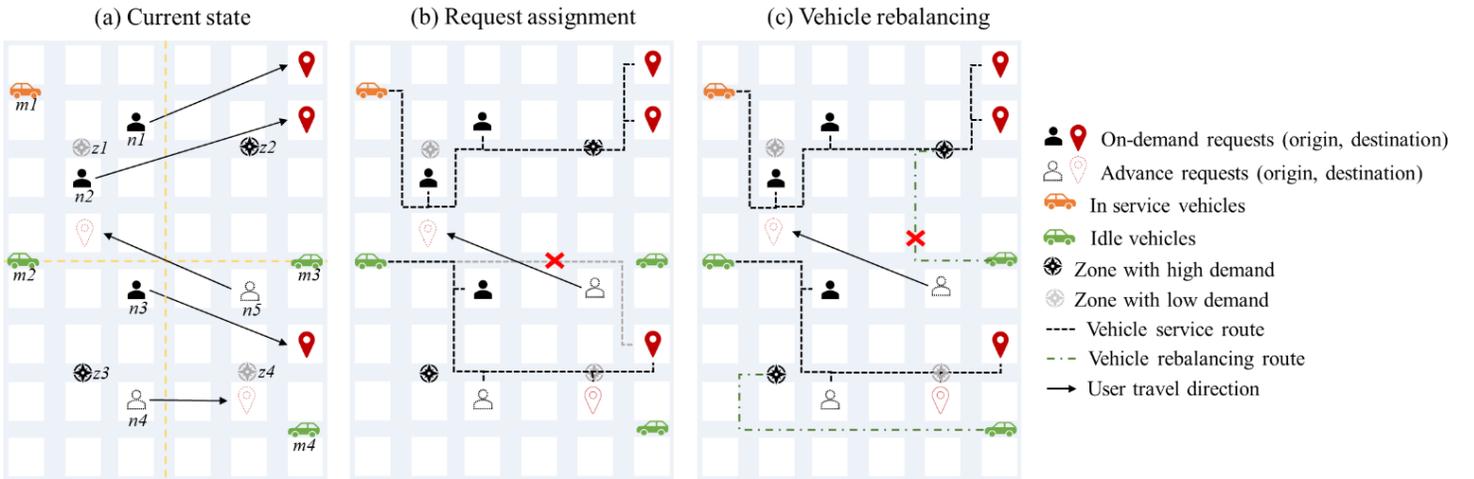

**Fig. 3.** Overview of the proposed approach for ride-pooling with advance requests. (a) current state includes the states of vehicles, requests, and zonal request rates; (b) Request assignment matches requests to vehicles and routes vehicles considering advance requests. For example, $m_1$ picks up $n_2$ and then $n_1$. $m_2$ picks up $n_3$ and is routed along the black-dashed path rather than the grey-dashed path given the advance request $n_4$; (c) vehicle rebalancing repositions an idle vehicle to a zone with a high expected request rate. For example, $m_4$ is repositioned to $z_3$ and $m_3$ stays at the current location given advance request $n_5$.

---

[2] Note that the notation $n$ represents a request and the subscript/superscript $(n, n + N)$ represents the (pick-up, drop-off) nodes of request $n$.

Requests can only be served by a compatible vehicle (in service, idle, waiting, and rebalancing); otherwise, they are not served in the current period or rejected. Vehicles can serve either single or shared trip requests (but not both at the same time). A vehicle can only have a single user on board if it commits to serve a solo ride request and may serve single or multiple users after the commitment.

We assume that once a request is picked up by a vehicle, the request will be delivered by that vehicle. A vehicle $m \in \mathcal{M}$ has a status $\delta_k^m$ at time $t_k$, where $\delta_k^m$ can be in service (serving a solo or shared trip requests), idle waiting (after in service or rebalancing), and idle rebalancing. A vehicle, except with a status of in-service serving a solo trip, is eligible to take new requests. For example, an in-service-shared vehicle may detour to pick up a new compatible request as long as the delays associated with the requests already in the vehicle do not become infeasible because of the revised drop-off times. The status indicator $\delta_k^m$ also specifies the priority of a vehicle to be assigned a request. For example, an idle vehicle has the highest priority to be used if the VMT difference between assigning a request to an in-service vehicle and the idle vehicle is within a prespecified range $\epsilon$, e.g. 1 mile.

The rebalancing decisions for idle vehicles are made every $\Delta t$ seconds (decision epoch). An idle vehicle can be rebalanced to another zone or remain at the current zone. However, an idle-after-rebalancing vehicle is not repositioned before it serves a new request at its current zone, or it has been idle for more than a given time threshold $\Psi$ (e.g., 5 minutes). Such a vehicle will stay at the zone where it is rebalanced to. An idle-in-rebalancing vehicle can terminate its rebalancing route before arriving at the destination and pick up new requests.

Given that the main interest in this paper is to investigate the potential of trip sharing and advance requests on reducing the impact of TNC operations on congestion and the environment, the objective of the assignment decisions is to minimize the expected VMT to serve as many requests as possible given constraints on requests (service types), LOS preferences (max delay and wait times) and operations (vehicle eligibility of service types, waiting, priority). However, the framework is flexible and can adopt other objectives, e.g. maximizing LOS.

The objective of vehicle rebalancing is to reposition idle vehicles over the operating area such that the likelihood of serving future requests is maximized, taking into consideration future expected demand and supply (available vehicles) in the various zones, while the rebalancing cost is minimized.

*A. Current State*

Let $\mathcal{N}_k$ be the set of active requests at epoch $k$. $\mathcal{N}_k$ includes new on-demand/advance requests and/or outstanding requests that were not served at epoch $k-1$. The state of request $n \in \mathcal{N}_k$ is represented by $(\alpha_k^n, \beta_k^n)$, where $\alpha_k^n$ and $\beta_k^n$ are the scheduled pick-up and drop-off times of request $n$ at epoch $k$.

Let $(v_k^m, \delta_k^m, S_k^m, T_k^m, O_k^m)$ be the state of vehicle $m \in \mathcal{M}$ at epoch $k$. $v_k^m, \delta_k^m, S_k^m, T_k^m, O_k^m$ are the location, service status, scheduled service stop sequence, times arriving at service stops, and occupancy of vehicle $m$ at epoch $k$. Figure 4 illustrates the work schedule of a vehicle. A schedule-block is a consecutive period of 'active' vehicle time between two consecutive periods of idle status (either idle waiting or rebalancing). A schedule-block starts with a pick-up and ends with a drop-off. Associated with a schedule block is a sequence of service stops $S_k^m$ (pick-up or drop-off) and the times $T_k^m$ when each stop is scheduled to take place. For example, the schedule block is denoted as $S_k^m = [2, n, N+n, N+2]$ and $T_k^m = [t_k^2, t_k^n, t_k^{N+n}, t_k^{N+2}]$. That is, pick up 2 at $t_k^2$, pick up $n$ at $t_k^n$, drop off $n$ at $t_k^{N+n}$, and drop off 2 at $t_k^{N+2}$. $S_k^m$ and $T_k^m$ are updated if new requests are assigned to vehicle $m$ or vehicle $m$ has finished serving a stop. Occupancy $O_k^m = [o_k^2, o_k^n, o_k^{N+n}, o_k^{N+2}]$ is the number of users in vehicle $m$ after serving each scheduled stop in $S_k^m$.

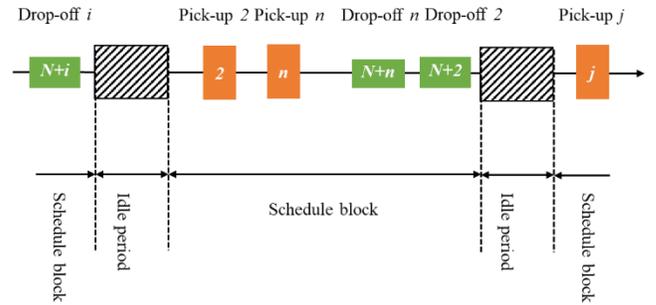

**Fig. 4.** Work schedule of a vehicle (adapted from [24]). Note that a stop $i$ is a pick-up node if $i \leq N$; drop-off node otherwise.

Let $\mathbb{N}_k = [(\alpha_k^n, \beta_k^n), \forall n \in \mathcal{N}_k]$ be the request state, and $\mathbb{M}_k = [(v_k^m, \gamma_k^m, S_k^m, T_k^m, O_k^m), \forall m \in \mathcal{M}]$ the vehicle state. The system state $x_k$ at epoch $k$ is defined as $x_k = (\mathbb{N}_k, \mathbb{M}_k)$.

*B. Assignment of Requests*

The assignment action $\mathcal{A}(x_k)$ involves identifying which requests should be served by which vehicles. The assignment problem is defined as:

*Given a set of requests $\mathcal{N}_k$ and fleet $\mathcal{M}$ at decision epoch $k$, find an optimal assignment of requests to vehicles $\mathcal{A}_k$ that minimizes the total service costs $C_k$ subject to constraints.*

The decision support platform (Figure 2) can deploy any algorithm for solving the assignment problem. The DRT and MOD literature includes a plethora of algorithms for solutions to the routing and assignment problem. In almost all cases, the service can be represented fundamentally as a 'dial-a-ride' (DAR) system [24-26], where requests must be served within a pick-up and/or drop-off time window.

These problems are computationally difficult to solve to optimality, and heuristics are typically used for their solution. Many heuristics have been proposed for solving the basic DAR problem and its extensions [25, 27], mostly inspired by the insertion heuristic proposed by Jaw, et al. [24]. Bertsimas et. al. [15] proposed an optimization framework coupled with a backbone algorithm to solve large-scale taxi routing problems with single trips. Recently, in the context of the on-demand ride-pooling problem, Alonso-Mora, et al. [10] proposed an anytime 'optimal' algorithm for matching and vehicle rebalancing. They use the concept of a shareability network [9] and an integer linear program formulation to solve the problem. While the algorithm identifies 'any-time optimal' solutions given sufficient time, its computational complexity increases exponentially with the capacity of the vehicles. Therefore it may not scale well for large problems and general vehicle capacities. There is also extensive work on autonomous MoD services with centrally controlled vehicles to improve LOS (e.g. minimizing waiting, etc.). These approaches attempt to find optimal vehicle routing and rebalancing strategies given real-time requests, using fluid approximation [28], queuing theory [29], Markov models [30], and model predictive control [19, 31] approaches. However, most consider single or low capacity vehicles and do not address well the ride-pooling problem.

Since the goal of the paper is to systematically evaluate ARRP services under various combinations of operational strategies and user characteristics, the number of scenarios that are generated from the combinations of the design parameters and user preferences is very large. Using efficient algorithms for the routing and scheduling phase is, therefore, essential from a computational point of view. While the algorithm proposed in [10] is an exciting development, it requires sufficient time to get optimal solutions, and it still has a computational complexity $\mathcal{O}(MN^\Omega)$, where $M$ is the number of vehicles, $N$ the number of requests, and $\Omega$ the capacity of the vehicles. It should be mentioned that for the solution of the taxi problem in Manhattan, the authors actually deployed a heuristic in order to deal with the computational issues of the original algorithm [10]. For example, if the number of passengers in the vehicle is larger than four, each additional request is assigned by only checking the routes that maintain the drop-off order of the passengers currently in the vehicle. Although heuristics do not guarantee optimality, they are useful approaches to find good solutions and evaluate and compare alternative operational designs [21-23].

For the purposes of this study, we developed an insertion heuristic (Algorithm 2) to identify good assignments based on [24] adapted to the characteristics of our ARRP problem. It processes requests sequentially, inserting one request at a time into the work schedule of a vehicle until all requests have been processed. It takes as inputs requests $\mathcal{N}_k$, vehicles $\mathcal{M}$, system state $x_k$, and model parameters, and outputs assignment action $\mathcal{A}_k$ at epoch $k$. The assignment action $\mathcal{A}_k$ is a $M$-dimensional vector where the $m^{th}$ element $a_k^m$ is the action associated with vehicle $m$ in $\mathcal{M}$. A vehicle action $a_k^m$ represents the requests in $\mathcal{N}_k$ to be served by vehicle $m$ and it may be an empty set.

**Algorithm 2** (RequestAssignment)

Input: $\mathcal{N}_k, \mathcal{M}, x_k, \epsilon$
Output: $\mathcal{A}_k$
0: **Initialize** $\mathcal{A}_k = \{a_k^m | a_k^m = \emptyset, \forall m \in \mathcal{M}\}; C = \{\ \}$
1: **For** $n$ in $AscendingSorted(\mathcal{N}_k)$:
2:      **If** $t_k > t_n^l$, **Reject** $n$, **Go to 1**
3:      **For** $m \in \mathcal{M}$:
4:          $IP = FeasibleInsertionPlan(n, m, x_k)$
5:          $C \leftarrow BestInsertionPlan(IP)$
6:      **If** $C$ is not $\emptyset$:
7:          **Find** $m0$ such that $C(m0) \leq C(m), \forall m \in \mathcal{M}_k^{serv}$
8:          **Find** $m1$ such that $C(m1) \leq C(m), \forall m \in \mathcal{M}_k^{idle}$
9:          **If** $|C(m1) - C(m0)| \leq \epsilon, a_k^{m1} \leftarrow a_k^{m1} \cup n$, **Go to 1**
10:         **Else**: $a_k^{m0} \leftarrow a_k^{m0} \cup n$, **Go to 1**
11:      **Else**: $\mathcal{N}_{k+1}^0 \leftarrow n$, **Go to 1**

Algorithm 2 begins by sorting requests in ascending order based on their earliest pickup time $t_n^e$. If the system time exceeds the latest pick-up time of request $n$, the request is rejected (line 2). The processing of a request $n$ proceeds as follows: (a) For vehicle $m$ in the set $\mathcal{M}$, the function *FeasibleInsertionPlan* finds all the feasible ways in which request $n$ can be inserted into the work schedule of vehicle $m$ (line 4). Insertion of a request is feasible only if it does not violate any constraints (as discussed below) for the new request and for all other requests already assigned to the vehicle. The idle vehicles are candidates to serve the new request, and the feasibility is determined by the request pickup time window. Function *BestInsertionPlan* finds the best plan of inserting the request $n$ to vehicle $m$ that results in a minimum additional cost and returns the cost (line 5). (b) After exploring all vehicles in the set $\mathcal{M}$, the final 'best' assignment decision is made by trading-off the cost and vehicle priority. The request is assigned to the vehicle with the highest priority if the cost difference between assigning to that vehicle and the vehicle with the lowest cost is within a predefined threshold $\epsilon$ (line 9); otherwise, the request is assigned to the vehicle with the least cost (line 10). (c) If the request is not assigned, it is added into $\mathcal{N}_{k+1}^0$ to be considered for assignment in the next epoch (line 11).

A feasible assignment of a request to a vehicle respects the user constraints (service type and LOS) and vehicle operations (available capacity, wait time). Insertion of a request to a vehicle schedule is feasible if:
- It does not violate vehicle capacity and LOS constraints for the newly assigned request and for all other requests already assigned to that vehicle.
- It respects vehicle status constraints of serving either single/shared requests, pick-up wait time (arrive early).

The set of feasible assignment action $a_k^m$ for vehicle $m$ is:

$$\begin{aligned}
\{a_k^m \subseteq \mathcal{N}_k | & \\
\cap_{m \in \mathcal{M}_k} a_k^m = \emptyset, & \quad (a) \\
t_s^e \leq t_k^s \leq t_s^l, \ \forall s \in \{s' | s' \in S_k^m, s' \leq N\}, & \quad (b) \\
(t_k^s - t_k^{s-N}) - \tau_{s-N,s} \leq D_{s-N}, \ \forall s \in \{s' | s' \in S_k^m, s' > N\}, & \quad (c) \\
o_k^s \leq \Omega_m, \ \forall s \in S_k^m, & \quad (d) \\
a_k^m = \emptyset, \ \forall m \in \mathcal{M}_k^{serv\_solo}, & \quad (e) \\
t_s^e - t_k^s \leq W_m, \ \forall s \in \{s' | s' \in S_k^m, s' \leq N\} & \quad (f) \\
\}
\end{aligned} \quad (1)$$

where $\mathcal{M}_k^{serv\_solo}$ is the set of vehicles serving solo trips at epoch $k$. Condition (a) states that each request is assigned to, at most, one vehicle. It is possible that some requests cannot be assigned to any vehicle at the current epoch time. Condition (b) requires that the pick-up occurs at or after time $t_s^e$, but no later than the time $t_s^l$ for any pick-up stop $s$ in schedule $S_k^m$. Condition (c) requires that the trip delay is no more than the acceptable maximum trip delay $D_{s-N}$ (request specific) for any drop-off stop $s$ in the vehicle schedule $S_k^m$. Condition (d) ensures that the number of passengers in a vehicle does not exceed its capacity for any stop in the vehicle schedule $S_k^m$. Condition (e) states that if a vehicle is serving a solo trip request, no additional requests can be assigned to that vehicle. Condition (f) guarantees that the vehicle wait time does not exceed the maximum allowed wait time $W_m$ for any pick-up stop $s$ in the vehicle schedule $S_k^m$.

The critical and most time-consuming step of Algorithm 2 is the *FeasibleInsertionPlan* function. It systematically examines all possible schedule sequences associated with every schedule block of a vehicle. The maximum number of possible insertions is $(|S| + 2)(|S| + 1)/2$, where $S$ is the schedule stop sequence and $|S|$ is the size of the sequence (number of stops). The worst-case complexity of exploring feasible insertion plans of a request is $\mathcal{O}(M \times \Omega^2)$, where $M$ is the number of vehicles and $\Omega$ vehicle capacity. Insertion explorations are independent for different vehicles and can be easily parallelized to improve computational efficiency. A fast screening method is developed that is efficient in checking feasible insertions and facilitating the solution of large-scale problems. It is based on the concepts of the service time window and maximum allowable delay $\theta$ for each stop (i.e. pick-up or drop-off) of the schedule block.

**Serving time window:** The required time window for a pick-up stop is the request pick-up time window $[t_n^e, t_n^l]$. For a drop-off stop, the required time window $[t_{n+N}^e, t_{n+N}^l]$ is estimated as:

$$t_{n+N}^e = t_n^e + \tau_{n,n+N}$$
$$t_{n+N}^l = t_n^l + \tau_{n,n+N} + D_n \quad (2)$$

For any request $n$, the schedule should satisfy:

$$t_n^e \leq \alpha_k^n \leq t_n^l$$
$$t_{n+N}^e \leq \beta_k^n \leq t_{n+N}^l \quad (3)$$

**Maximum allowable delay:** Let $j$ denote a stop index in a schedule block. The maximum allowable time $\theta_k^j$ at stop $j$ is determined by the minimum of the maximum amount of time by which the visit at every following stop (not including stop $j-1$) can be delayed due to the insertion of a new request.

$$\theta_k^j = \min_{j \leq j' \leq |S|} \left( t_{s(j')}^l - t_k^{s(j)} \right) \quad (4)$$

Figure 5 shows an example of possible insertions. Each stop of the schedule block has attributes $(\theta^j, e^j, \ell^j, t^j)$, representing the maximum allowable delay, the required time window, and the scheduled serving time. Insertion plan B attempts to insert the pick-up stop $n$ between the 2nd and 3rd stops and the drop-off between the 3rd and 4th stops in the schedule block with 4 stops. The additional delayed time at stop $j = 3$ due to picking up $n$ is $\sigma = \tau_{2,n} + \tau_{n,N+2} - \tau_{2,N+2}$. If $\sigma \leq \theta_k^3$, then it is feasible to insert pick-up stop $n$ at the designated location without violating time windows for requests already in the schedule. Similarly, the drop-off insertion feasibility is checked after updating $\theta_k^j, \forall j$.

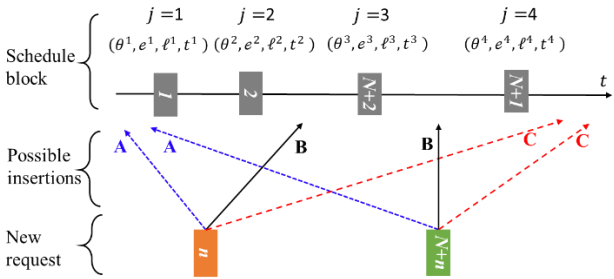

**Fig. 5.** Possible insertions. (A) Both the pick-up and drop-off stops are inserted between two consecutive stops; (B) The pick-up and drop-off stops are separated by one or more stops; (C) Both the pick-up and drop-off stops are at the end of the current schedule.

The fast screening improves the insertion efficiency: 1) avoids updating/checking detailed stop attributes in Equation 6 over all stops by only checking the maximum allowable delay at one stop; 2) reduces possible insertions. The insertion is more efficient by examining the possible insertions from the last to the first stop, as it can stop exploring the preceding stops whenever a pick-up insertion violates the maximum allowable delay at stop $j$.

### C. Vehicle Rebalancing

The rebalancing action $\mathcal{B}(x_k)$ specifies which zone idle vehicles should be repositioned to. The problem is defined as:

*Given a fleet $\mathcal{M}$ and zonal request rates $\Lambda$ at decision epoch $k$, reposition the eligible idle vehicles to zones $\mathcal{B}(x_k)$ to maximize the likelihood of serving a request in the future while the rebalancing cost is minimized.*

Rebalancing vehicles travel along the shortest path. The vehicle can pick up requests en-route and terminate rebalancing. A rebalancing vehicle upon reaching its destination can pick up a request, but if none is available cannot be rebalanced again for another time $\Psi$ (e.g., 5 minutes). Maximizing the likelihood of all individual vehicles having at least one request to serve at their destinations is based on the probability that the assigned zones have at least $r_z$ requests. $r_z$ is the number of vehicles assigned to zone z for rebalancing.

$$\mathcal{J}_{service} = \prod_z P_z \quad (5)$$

where, $P_z = prob(r \geq r_z)$ is the probability that a zone has at least $r_z$ requests for the future period (e.g., next 15 minutes), and $r_z = \sum_m y_{mz}$. $y_{mz}$ is a binary variable, $y_{mz} = 1$ if vehicle $m$ is repositioned to zone $z$; 0 otherwise.

Rebalancing cost is defined as the total VMT of rebalancing.

$$\mathcal{J}_{cost} = \sum_m \sum_z (y_{mz} u_{mz}) \quad (6)$$

where, $u_{mz}$ is the travel distance for vehicle $m$ rebalancing to zone $z$ from its current location.

The rebalancing problem is formulated as a bi-objective optimization problem (maximize service and minimize cost):

$$\begin{aligned}
&\text{Maximize} && \mathcal{J}_{service} \\
&\text{Minimize} && \mathcal{J}_{cost} \\
&s.t. \\
& && \sum_z (y_{mz} u_{mz}) \leq \Phi, \forall m \in \mathcal{M} && (a) \\
& && \sum_z y_{mz} = 1, \forall m \in \mathcal{M} && (b) \\
& && y_{mz} \in \{0,1\}, \forall m \in \mathcal{M}, \forall z \in \mathcal{Z} && (c)
\end{aligned} \quad (7)$$

where, $\Phi$ is the maximum rebalancing distance. Condition (a) states that the feasible candidate rebalancing zones for a vehicle are within distance $\Phi$ from the current location. Constraints (b) and (c) guarantee that a vehicle is assigned exactly to one zone, including the zone that a vehicle is currently located at. More than one vehicle can be assigned to the same zone.

For the multiple-criteria optimization problem in Equation (7), a single solution simultaneously optimizing each objective may not exist, but rather a number of Pareto optimal solutions [32]. A heuristic algorithm is proposed to find a good solution for the rebalancing problem. It takes as inputs vehicles $\mathcal{M}$, system state $x_k$, and model parameters, and outputs rebalancing action $\mathcal{B}_k$ at epoch $k$. The action $\mathcal{B}_k$ is a $M$-dimensional vector where the $m^{th}$ element $b_k^m$ is the action associated with vehicle $m$. A vehicle action $b_k^m$ represents the zone in $\mathcal{Z}$ that vehicle $m$ is rebalanced to (including the current location).

**Algorithm 3** (RebalanceVehicles)

---
**Inputs**: $\mathcal{M}, x_k, \Lambda, \Psi, \Phi$
**Output**: $\mathcal{B}_k$
1: **Initialize** $\mathcal{B}_k = \{b_k^m | b_k^m = \emptyset, \forall m \in \mathcal{M}\}; r_z = 1, \forall z \in \mathcal{Z}$
2: $\mathcal{M}_k^{reb} = EligibleRebVehicles(\mathcal{M}, x_k, \Psi)$
3: $\mathcal{Z}_k = AccessibleZones(\mathcal{M}_k^{reb}, x_k, \Phi)$
4: **While** $\mathcal{M}_k^{reb}$ is not $\emptyset$:
5: $\quad P = \{P_z | P_z = prob(r \geq r_z | \Lambda), \forall z \in \mathcal{Z}_k\}$
6: $\quad$ **For** $P_z, z$ in $DescendingSorted(\mathbf{P})$:
7: $\quad\quad \mathcal{M}_{cand} = CandidateVehicles(\mathcal{M}_k^{reb}, z, \Phi)$
8: $\quad\quad$ **If** $\mathcal{M}_{cand}$ is not $\emptyset$:
9: $\quad\quad\quad$ **Find** $m^*$ such that $u_{m^*z} \leq u_{mz}, \forall m \in \mathcal{M}_{cand}$
10: $\quad\quad\quad b_k^{m^*} = z, \mathcal{M}_k^{reb} = \mathcal{M}_k^{reb} \setminus m^*$
11: $\quad\quad\quad r_z = r_z + 1$, **Go to 4**
12: $\quad\quad$ **Else**: **Go to 6**

---

The algorithm first screens vehicles eligible for rebalancing (line 2, i.e. wait vehicles after service and rebalanced vehicles after waiting for a certain time, e.g. 5 minutes), finds the set of zones $\mathcal{Z}_k$ accessible by any vehicle in $\mathcal{M}_k^{reb}$ given the maximum rebalancing distance $\Phi$ (line 3), and calculates the probability $P_z$ of having at least $r_z$ requests for any zone in $\mathcal{Z}_k$ (line 5)[3]. It then examines zones in descending order of probability $\mathbf{P}$ (line 6), assigns the nearest eligible vehicle to the zone with the highest probability, pops up the assigned vehicle, and increases the number of rebalancing vehicles to that zone $r_z$ by 1 (lines 8-11). Line 7 finds the set of vehicles $\mathcal{M}_{cand}$ whose travel distance to zone $z$ is within maximum distance $\Phi$. No candidate vehicle may exist for a zone, even if it has a high request rate (line 12). The algorithm terminates when all idle vehicles are assigned to a rebalancing zone (including their current location) (line 4).

Note that Algorithm 3 can accommodate various rebalancing strategies by using different considerations to calculate $P_z$ (line 5). By using the expected request rate $\lambda_{zk}$, it targets the zones with high expected demands to rebalance vehicles to. If $\lambda_{zk} = 1$ for all zones and epochs, it performs uniform rebalancing, where all zones have an equal probability of being assigned a vehicle. Expected future requests (demand) and the number of vehicles expected to arrive at the zone based on their current assignments (supply) may also be considered simultaneously, resulting in an "effective" probability. It can be defined, for example, as $P_z = prob(r \geq r_z + v_z)$, where $v_z$ is the expected number of vehicles in service destined to that zone and they are expected to arrive during the same period.

*D. Update System State*

Given assignment and rebalancing actions $\mathcal{A}(x_k)$ and $\mathcal{B}(x_k)$, the system state at epoch $k + 1$ is updated to $x_{k+1}$. A request $n$ transitions out of the set $\mathcal{N}_k$ when it is picked up or rejected. A request is rejected if it is no longer feasible ($t_k > t_n^l$). Requests that remain in $\mathcal{N}_k$ are rolled over to $\mathcal{N}_{k+1}^0$ at epoch $k + 1$ along with all new requests.

If the action $a_k^m$ for vehicle $m \in \mathcal{M}$ at epoch $k$ contains new requests, its set of stops $S_k^m$ is updated by inserting the pick-up and drop-off stops of new requests. Let $s(j) = S_k^m[j]$ be the stop with index $j$ in the scheduled service stop sequence. For the example in Figure 3, $s(1)$ is stop 2. The time $t_k^s$ to serve each stop $s \in S_k^m$ is updated as:

$$t_k^{s(j)} = \begin{cases} max\left(t_k^{s(j-1)} + \tau_{s(j-1),s(j)}, t_{s(j)}^e\right), & if\ s(j) \leq N \\ t_k^{s(j-1)} + \tau_{s(j-1),s(j)}, & if\ s(j) > N \end{cases} \quad (8)$$

where $s(0) = v_k^m$, the current vehicle location.

A request is always served by the vehicle originally assigned to, although the service order of the requests may change at following epochs if a better assignment is found, either for onboard passengers or not yet picked up users. Eligible vehicles are rebalanced to zones following the shortest (travel time) path, and vehicle states are updated accordingly.

The vehicle state $\mathbb{M}_{k+1}$ is updated for vehicles in service and rebalancing if any scheduled stop in $S_k^m$ is executed. Also, the request state $\mathbb{N}_k$ is updated accordingly.

*E. Cost*

Let $\mathcal{M}_k^a = \{m | m \in \mathcal{M}, a_k^m \neq \emptyset\}$ be the set of vehicles being assigned new requests at epoch $k$, and $S_k^{m,a}$ the schedule of vehicle $m$ given action $a_k^m$. The corresponding cost function is:

$$C_k(\mathcal{A}(x_k)) = \sum_{m \in \mathcal{M}_k^a} \left( \sum_{j=1}^{|S_k^{m,a}|} u_{s(j-1),s(j)} - \sum_{j=1}^{|S_k^m|} u_{s(j-1),s(j)} \right) \quad (9)$$

where $|S_k^m|$ is the number of elements (stops) in $S_k^m$.

The cost associated with assignment action $\mathcal{A}(x_k)$ is the sum of the additional VMT to serve the new requests in $\mathcal{N}_k$ over all vehicles in $\mathcal{M}_k^a$. The additional VMT is the distance difference between the schedule $S_k^{m,a}$ under action $a_k^m$ and $S_k^m$.

## IV. EXPERIMENTS AND RESULTS

The case study focuses on exploring the ARRP performance under different operating characteristics, service preferences, and traffic conditions.

A large-scale TNC open-source dataset [4] from Chengdu, China, provided by DiDi Chuxing, is used. For each trip, the dataset contains a request ID, vehicle ID, and pick-up and drop-off times and locations. Figure 6 shows the heatmap of the distribution of pick-up and drop-off locations for a morning period (7:30-8:30 am). The pick-up locations are scattered across the network, while drop-offs are relatively concentrated.

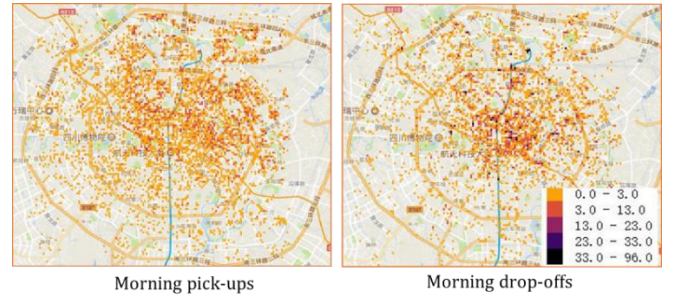

**Fig. 6.** Distribution of pick-up and drop-off locations for a morning peak period 7:30-8:30 am (cell area 150m*150m)

The dataset does not contain information about ride pooling. By linking actual rides (origin/destination, timestamps) with vehicle trajectory data, the shared trips can be inferred. Figure 7 shows that shared trips account for 3% to 8% of all trips. No shared service is offered from 0:00 to 6:00 for safety reasons. Figure 7 also shows the shared service hours ratio, defined as the ratio of the total shared service hours (sum of the shared hours * the number of shared riders) over the total system service hours (sum of the service hours * vehicle seats). The shared service hours ratio is around 2%, equivalent to 1.02 riders per trip. The number of customers requesting a shared service is not known. Due to the lack of more detailed information, it is unclear what causes the relatively low number of shared trips, e.g. lack of interest in sharing, spatiotemporal characteristics of requests, and/or operating practices.

---

[3] Note that $P_z = 1$ if there is outstanding request $\mathcal{N}_k^0$ in that zone; otherwise $P_z = prob(r \geq r_z | \Lambda)$. The number of outstanding requests of a zone is reduced by 1 when an idle vehicle is assigned to that zone till it becomes 0.

[4] https://outreach.didichuxing.com/research/opendata/en/

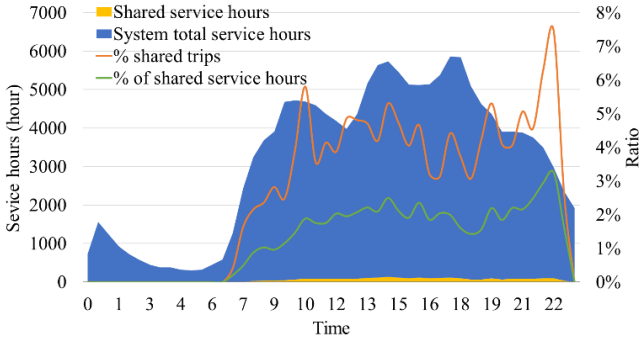

**Fig. 7.** Distribution of shared rides by time of day

To assess the impact of various factors on the performance of the ARRP services, we use the actual rides as requests. The rides within the 3rd ring of the city (20km *20km area, covering 85% of requests) are considered. For vehicle initialization and rebalancing purposes, we partition the area into zones of 1 km * 1 km. The average request rate in each zone is calculated in 15-minute intervals using the overall weekday data. The road network is extracted from Open Street Map (OSM), including nodes, links, and link attributes (length and road types) [33]. The free-flow travel time is estimated from the link length and link travel speed limits based on the road type. The travel times at different times of the day are approximated as the free-flow travel time multiplied by appropriate factors.

Vehicles are initialized in two steps. They are randomly assigned to a zone proportionally to the zonal request rate $\lambda_z$, and randomly positioned to one of the network nodes in the corresponding zone.

Rebalancing vehicles are positioned to the zone centroid. The Floyd-Warshall algorithm is used to calculate the shortest paths for all node pairs [34]. After experimenting, the following parameters are used for Algorithm 3: the rebalancing strategy considers outstanding requests and the expected future demand and supply. Rebalancing vehicles cannot be rebalanced for another $\Psi = 5\ minutes$ after they reach their destination. The maximum rebalancing distance is $\Phi = 5\ km$. The tradeoff between vehicle priority and VMT is set to $\epsilon = 1\ km$.

Current operating practices (on-demand) are represented by a scenario with no advance requests and decision epoch $\Delta t = 30$ seconds. The operations are reactive, scheduling just the on-demand requests received during a short time window. For the model with advance requests, the vehicle schedule is optimized every decision epoch, e.g. every 30 seconds, given vehicle states and the on-demand and advance requests.

A set of performance metrics with respect to sustainability, LOS, and fleet utilization is used to evaluate various designs:

**Sustainability**
- *Vehicle miles per request (VMR)* is the average VMT per served request. It includes VMT for picking up and serving requests and rebalancing idle vehicles.
- *VMR_Service* is the average VMT (per served request) while serving requests.
- *VMR_Idle* is the average VMT (per served request) while picking up requests and rebalancing idle vehicles.
- *% shared trips* is the percentage of all served requests that are shared.

**LOS**
- *Average wait time* over all the served requests.
- *Average trip delay* (deviation from direct service) over all served requests.
- *% served requests* is the percentage of all requests actually served.

**Fleet Utilization**
- *# of active vehicles* is the total number of vehicles that served at least one request in the period of analysis.
- *Maximum vehicle occupancy* is the maximum number of customers sharing a vehicle at any time during its trip.

*A. Overall Potential*

Figure 8 compares the VMR of three operating models: a) all users are willing to share a ride and place requests in advance with a 30-minute horizon (Share_ADV30); b) all customers are willing to share but no requests in advance (Share_ADV0); and c) no sharing and no requests in advance (Base_Case). In all cases, requests are assigned every 30 seconds, LOS constraints with maximum wait 7 minutes and maximum delay 15 minutes (referred to as neutral LOS), vehicle capacity of 4 passengers per vehicle, and average congestion (normal traffic conditions), fleet size of 3,000 vehicles (enough to serve all requests for the three operating models). All requests for the period from 6:00 to 23:00 are considered (125,320 requests).

Introducing advance requests consistently improves system performance in terms of VMR. In the base case, where all requests are for solo trips, 639,170 miles are required to serve the 125,320 requests for the day. Compared to the base case, a system with all users willing to share but with no advance requests will reduce VMT by 242,448 miles (37.9%), while a system with all passengers willing to share and place requests in advance and a horizon of 30 min will save a total of 328,545 miles (51.4%) of the base case VMT. These results represent significant savings in terms of environmental costs as well since these are directly related to VMT.

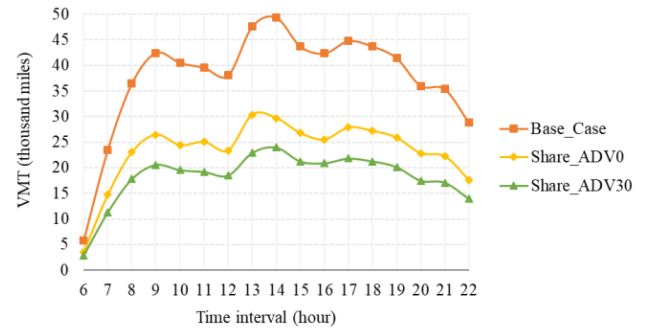

**Fig. 8.** Comparison of VMT by time of day for different operating models: the base case with no shared trips and no requests in advance, sharing but no advance requests, and advance requests with $H = 30\ min$ and sharing. Other settings: vehicle capacity 4, neutral LOS constraints.

In terms of the LOS, the no advance requests system with all willing to share has an average wait time of 4.2 minutes and trip delay of 8.5 minutes, while the system with requests in advance has an average wait time of 3.3 minutes and trip delay of 8.8 minutes. Using requests in advance not only reduces the wait time compared to the case of no advance requests and shared trips but, interestingly, it results in wait time even lower than the base case as well (when all trips are single, the wait time is 3.6 minutes) despite the fact that trips are shared, most likely due to the improved scheduling and operating efficiency.

### B. Impact of Specific Factors and Tradeoffs

The above results show the potential of a ride pooling system with advance requests in reducing VMT while maintaining a competitive (or even better) LOS. While the advance requests service is promising, it is important to evaluate the sensitivity of its performance to a wide range of conditions in terms of operations, user preferences, and traffic conditions. Table II summarizes the various design aspects and parameter settings of the empirical analysis.

TABLE II
EXPERIMENTAL DESIGN AND PARAMETER SETTINGS

| Dimension | Parameter | Settings* |
|---|---|---|
| Operations | Fleet size | $M = 1500$ |
| | Vehicle capacity | $C \in [2, 4, 7, 10]$ passengers/veh |
| | Advance horizon | $H \in [0, 5, 15, 30, 60]$ minutes |
| User preferences | The fraction of users willing to share (WSF) | $f_s \in [0\%, 33\%, 67\%, 100\%]$ |
| | The fraction of users with requests in advance (ARF) | $f_a \in [0\%, 33\%, 67\%, 100\%]$ |
| | LOS constraints (wait, delay)* | Strict: (5min, 10min) Neutral (NEU): (7min, 15min) Flexible (FLE): (10min, 20min) |
| Traffic conditions | Speed^ | Low (congested) Medium (normal) High (light) |

* (maximum passenger wait time, maximum passenger trip delay time). ^ the low and medium speeds are 1/2 and 2/3 of the high speed, respectively.

The impact of the various factors listed in Table II on ride pooling performance is explored by operating the system under various scenarios. Each scenario is a combination of the parameters listed in Table II, resulting in 2,880 scenarios. The morning period (7:30-8:30 am) with 4,877 requests is used for the analysis. The system is operated using Algorithm 1 with a decision epoch of 30 seconds.

The fleet size used is an important consideration. Fleet size is impacted by all the factors discussed above, as well as the rebalancing strategies. We analyzed the minimum fleet size using a minimum cost flow formulation of the problem [35]. The estimated fleet size is a lower bound on the actual fleet size requirements. The required minimum fleet size to serve all requests in the case of solo/on-demand trips with neutral LOS constraints and normal traffic is 1,300 vehicles. A fleet of 1,500 is used for analysis allowing buffer considering rebalancing and uncertainties not accounted for by the problem representation in [35]. The same set of requests (with attributes) is used, and vehicles are initialized to the same locations in all scenarios.

Given a large number of scenarios (2,880), to facilitate the analysis and, four linear regression models are estimated. These models provide useful insight into the importance of various factors and their trade-offs. The dependent variables are VMR, average vehicle miles to serve a request (in miles); STF, actually shared trip fraction; WT, wait time (in minutes); and DT delay time (in minutes). The explanatory variables are related to the fractions of customers willing to share and place requests in advance, LOS constraints, and traffic conditions. Table III summarizes the regression results. Only significant variables are used in the reported model.

TABLE III
REGRESSION RESULTS FOR VMR, SHARED TRIPS, WAIT AND DELAY TIMES

| Variables# | VMR* | STF* | WT* | DT* |
|---|---|---|---|---|
| Constant | 5.66 (113) | -0.16 (-16.5) | 2.76 (50.8) | -1.31 (-11.4) |
| WSF | -1.12 (-14) | 0.50 (32.8) | 0.56 (7.03) | 1.85 (10.2) |
| NEU×WSF | -0.16 (-2.2) | 0.04 (2.9) | 0.18 (2.0) | 2.06 (8.9) |
| FLE×WSF | -0.18 (-2.0) | 0.06 (4.2) | 0.44 (4.3) | 4.5 (19.4) |
| H5×ARF | -0.40 (-6.7) | 0.04 (3.3) | -1.14 (-18.3) | |
| H15×ARF | -0.59 (-9.9) | 0.05 (3.7) | -1.43 (-23.0) | |
| H30×ARF | -0.67 (-11.0) | 0.05 (3.1) | -1.53 (-25.0) | -0.29 (-2.20) |
| H60×ARF | -0.68 (-11.0) | 0.06 (2.0) | -1.56 (-25.0) | -0.44 (-3.24) |
| NEU | 0.03 (0.59) | | 0.85 (13.4) | 0.25 (1.89) |
| FLE | 0.07 (1.46) | | 2.16 (34.0) | 0.30 (2.12) |
| Capacity | -0.10 (-5.6) | 0.08 (17.4) | 0.14 (7.14) | 1.19 (26.6) |
| Capacity^2 | 0.004 (3.8) | -0.005 (-10) | -0.013 (-6.7) | -0.07 (-16.9) |
| Normal | | 0.05 (5.2) | -0.26 (-6.9) | 0.16 (1.9) |
| Light | 0.43 (13.6) | 0.06 (6.6) | -0.48 (-12.5) | 0.64 (7.4) |
| Adj. R^2 | 0.42 | 0.72 | 0.65 | 0.74 |
| Observations | | 2,880 | | |

*: estimated coefficients (t-value in parenthesis).
#: WSF is the willingness to share fraction; ARF is the willingness to place requests in advance fraction; × indicates an interaction term (multiplication of two variables). NEU, FLE are neutral and flexible LOS constraints, the reference case is strict LOS; H5, H15, H30, H60 are advance request horizons of 5, 15, 30, 60 minutes, the reference case is no advance requests; Capacity equals to the actual vehicle capacity if WSF>0, otherwise 0; Normal, Light are normal and light traffic conditions, the reference case is congested traffic;

The estimation results indicate that VMR decreases as the users' willingness to share increases at the expense of higher wait time and trip delay since most of the trips are being shared. However, more users placing requests in advance and longer advance request horizons reduce not only VMR but also the wait time. They also increase the % of trips actually shared. Routing efficiency can be improved when more requests are known in advance and incorporated into the assignment and rebalancing decisions. For example, instead of rebalancing an idle vehicle immediately, knowledge of advance requests in the near future may result in vehicles waiting to serve those requests instead of rebalancing. Similarly, advance requests can help determine more efficiently which requests should be assigned to which vehicles and which requests should be delayed until later. As a result, operations with advance requests produce more efficient routes, thus reducing idle driving for pickup or rebalancing, eventually improving sustainability (VMR) and LOS (wait times). Given the same fraction of users requesting rides in advance, increasing the advance request horizon has a diminishing impact on VMR and wait times. It indicates that the system with advance requests could realize most of the benefits even with modest advance horizons (near-on-demand requests).

More flexibility in LOS constraints further decreases VMR, increases the % of shared trips, but also increases wait time and trip delay. The impact on VMR from increasing flexibility in LOS is diminishing (strict to neutral to flexible). Assuming that all users are willing to share, changing the LOS constraints from strict to neutral increases the wait time by 1.03 minutes and delay time by 2.31 minutes, and from neutral to flexible increases the wait time by 1.6 minutes and delay time by 2.5 minutes. Larger vehicle capacity reduces VMR and increases the actual % of shared trips but also increases wait time and trip delay for users. Higher capacity vehicles increase the sharing opportunities and thus decrease vehicle miles to serve requests, but increase user wait and delay times due to detouring for shared trips. In general, the impact is nonlinear, and the marginal contribution of an additional unit of capacity decreases as capacity increases. This is expected because the requests that can be actually shared is determined more by other factors when the seating availability is not a bottleneck.

Traffic conditions have interesting effects. VMR increases when traffic is light compared to congested traffic, but no significant impact is observed for normal traffic. Better traffic conditions increase the % of trips actually shared because the LOS constraints are more likely to be satisfied in terms of wait time and trip delay. However, they could also increase the possibility that idle vehicles drive longer distances for picking up or rebalancing, which could cancel out or outweigh the saved VMT from shared trips. The wait time decreases since vehicles can drive faster to pick up users under better traffic conditions. The delay time increases mainly due to more shared trips.

According to Table III, the base case (WSF = 0, ARF = 0, Normal = 1, NEU = 1) has a VMR of 5.59 miles per served request, wait time 3.48 minutes. Assuming all users are neutral, a 10% increase in the willingness to share fraction decreases the VMR by 0.13 miles (-2.28%) but increases wait time by 4.4 (+2.13%) seconds. A 10% increase in the willingness to request trips 30 minutes in advance can decrease the VMR by 0.067 miles (-1.20%) and wait time by 9.2 seconds (-4.40%). Comparatively, encouraging shared trips is more effective in reducing VMR, while incentivizing placing requests in advance can improve the LOS and provide additional benefits with reduced VMR. In addition, encouraging users to request trips in advance has more impact on LOS than improving traffic conditions. For example, improving traffic conditions from heavy to light reduces the average wait time by 28 seconds, while incentivizing 35% of users to place requests 30 minutes ahead of their planned departure saves 31 seconds.

From a user behavior point of view, willingness to share and place requests in advance are two important aspects that operators could target, using for example pricing strategies, to incentivize them towards more sustainable choices. Figure 9 shows the performance in terms of VMR (sustainability) and LOS as a function of the users' willingness to share and request service in advance. The system operates with a fleet size of 1,500 vehicles, vehicle capacity 4, neutral LOS constraints, advance requests horizon of 30 minutes, and normal traffic.

The results indicate that increasing the willingness to share and place requests in advance decreases VMR and rejected trips. Willingness to share is an important prerequisite to achieve the full benefits of advance requests. For example, with 100% willingness to share and 0% requests in advance, the VMR is 3.0 miles, while with 100% requests in advance the VMR is 2.5 miles (-16.7%). However, with 0% willingness to share and 100% requests in advance, the VMR is 5.1 miles (a decrease of 3.77%, compared to 5.3 miles with 0% requests in advance). Advance knowledge of requests is more useful when trips are shared, as the scheduling algorithms have more opportunity to match requests and effectively route vehicles. If all trips are single, the benefits from advance requests are much more limited. As the fraction of willingness to share decreases, requests in advance are important in reducing the number of rejected trips. Operations based on requests in advance are also more effective in reducing wait time for the passengers (mitigating the impact of shared trips on waiting) with no obvious impact on delays.

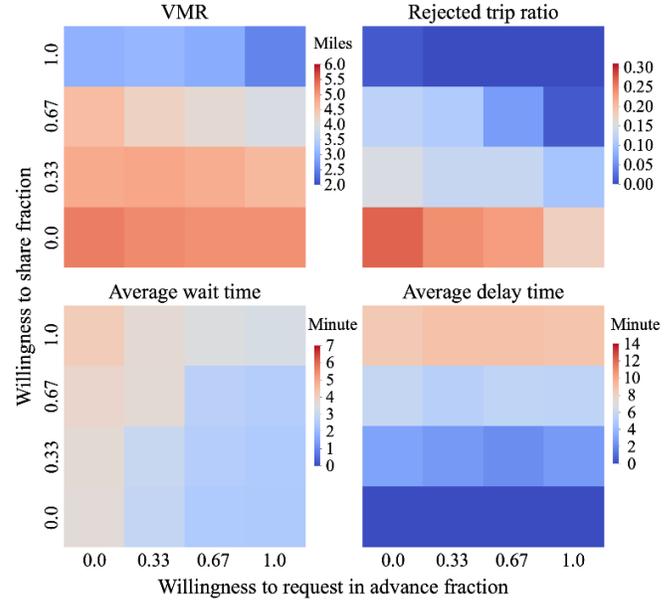

**Fig. 9.** Impact of willingness to share and place requests in advance on VMR, rejected trip ratio, average wait time, and average delay time.

### C. Detailed Analysis

The results in the previous section highlight the overall potential of the advance requests strategy. In this section, we discuss in more detail the performance of various operating strategies. The same set of 4,877 requests as in Section B is used. Unless otherwise specified, the discussion is based on the case with a fleet size of 1,500 vehicles, all users willing to share, neutral LOS constraints, and normal traffic conditions.

We use specific cases to systematically explore the operating performance of ride pooling systems with and without requests in advance in terms of sustainability, LOS, and fleet utilization. The performance is examined for a combination of factors that are expected to influence performance. From an operations planning perspective, two of the main design parameters are advance request horizon and vehicle capacity. LOS constraints influence system performance as they impact their flexibility.

*Sustainability.* Figure 10 shows how the advance requests horizon and vehicle capacity impact VMR and the % of trips actually shared. The values for VMR metrics are the ratios of the performance metric for a given scenario to that of the case with no advance requests. Operating with advance requests improves VMR compared to the case of no advance requests ($H = 0$), regardless of capacity. However, the magnitude of the impact varies for different capacities. The advance requests service reduces VMR_Service compared to the no advance requests case, but no obvious difference exists for the various advance request horizons when the vehicle capacity exceeds 4. The VMR_idle decreases consistently with the increase of the advance requests horizon, confirming that knowledge of future requests results in more efficient assignments and routing and can reduce idle driving for picking up new requests and rebalancing for future requests. Higher vehicle capacities improve VMR performance. However, there are diminishing returns when the capacity exceeds 7. Almost all trips are shared when the vehicle capacity exceeds 2, regardless of advance requests horizon. When the capacity is 2 (which may be of practical interest as services that limit the number of shared rides may be much more attractive to users), longer advance request horizons contribute significantly to increasing shared trips. When the capacity is 1 (no shared trips), advance requests improve VMR, however the benefits are more significant when trips are shared and vehicle capacity high enough to serve them.

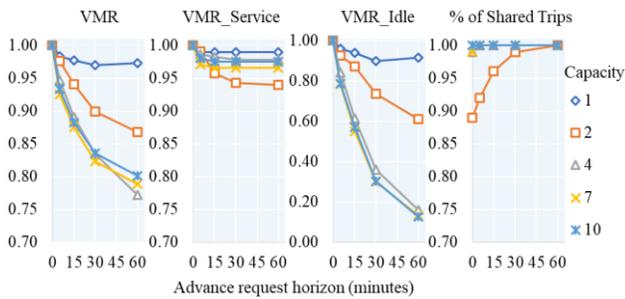

**Fig. 10.** Sustainability performance as a function of advance requests horizon and vehicle capacity. Each point (for VMR metrics) is the ratio of the corresponding metric to the reference case of no requests in advance.

Figure 11 compares the spatial distribution of the % of shared trips for the services with and without requests in advance. Each circle is a spatial cluster grouped based on the Euclidean distance between pick-up locations. The circle radius reflects the number of pick-ups within the cluster. The circle is color-coded given the attributes of interest, e.g. the fraction of shared trips within the cluster. Compared to the case with no advance requests, the case with advance requests has more shared trips, especially in areas away from the center.

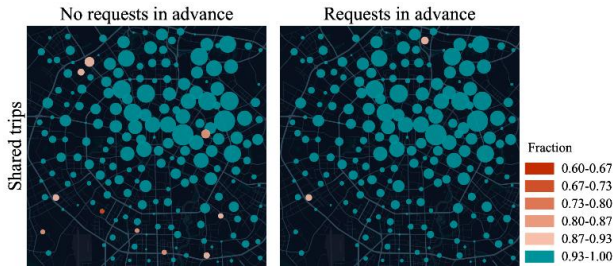

**Fig. 11.** Spatial distribution of the fraction of shared trips for the case with requests in advance (horizon 30 minutes) and no requests in advance. The vehicle capacity is 2.

*LOS*. Figure 12 shows the LOS performance with respect to LOS constraints, advance requests horizon, and vehicle capacity. In general, long advance requests horizons, high flexibility in LOS constraints, and high capacity increase the % of requests served. Almost all trips are served when the vehicle capacity exceeds 2, regardless of LOS constraints and advance requests horizons. For vehicle capacity of 2 and strict LOS constraints, almost all trips could be covered with a modest advance requests horizon of 30 minutes. If the horizon is 15 minutes more than 95% of the trips are served regardless of LOS constraints. When the capacity is 1 (solo trips), advance requests and flexible LOS constraints help increase the number of served trips.

The wait and delay times are significantly impacted by LOS constraints and capacity. Higher flexibility in LOS and higher capacity increase wait times and delay times. However, given the same LOS constraints and capacity, the use of advance requests helps moderate the impacts and reduces wait time. The length of the horizon has some impact, but most of the benefits are realized with $H = 15$. The additional impact from advance horizons longer than 15 minutes is only marginal. Trip length, demand density and patterns could be the potential reasons which determine the opportunity to match sharing requests. With average trip length less than 6 miles an advance requests horizon of 15 minutes is long enough to provide the needed information to take advantage of the available opportunities for pooling requests to the full extent in the analyzed dataset. In addition, the delay and wait times are not sensitive to capacity when it exceeds 4. The delay times (and to some extend wait times) are impacted by the number of requests that are actually shared (or the average occupancy of the vehicles). Figure 10 shows that almost all trips are shared when the vehicle capacity (with or without requests in advance). The reason for the insensitivity of LOS to the vehicle capacity when it is over 4 is two-fold. The LOS constraints do not allow passengers to experience delay and wait times above those limits; and the level of demand in the area does not support higher occupancy levels.

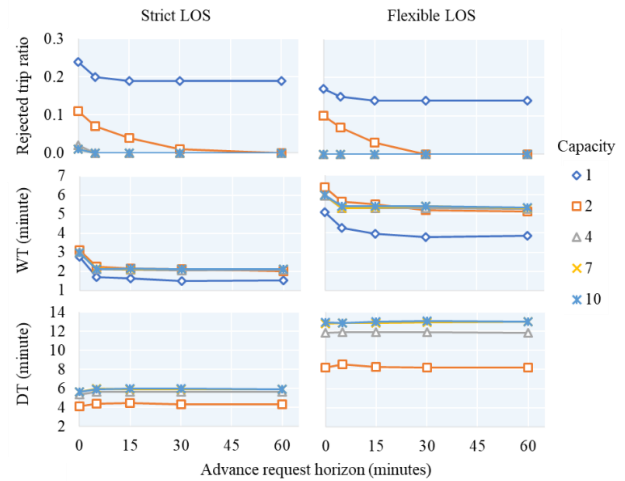

**Fig. 12.** LOS performance with respect to LOS constraints, advance requests horizon, and vehicle capacity in terms of the rejected trip fraction (top), average wait time (middle), and average delay time (bottom).

Figure 13 compares the spatial distribution of wait times and rejected trips for the services with and without requests in advance. Most of the trips with high wait times are concentrated in the city's central area (center of the map) due to the high probability of sharing trips given the high demand. The wait time tends to decrease further away from the central area. The service with advance requests not only has lower average wait times compared to the service with no advance requests but it also contributes to better service in areas away from the center. The system with no advance requests has more rejected trips than the system with advance requests. Many of the rejected trips are located in the northeast corner of the area, mainly because of the insufficient 'effective' supply (capacity actually utilized) in that area. On the other hand, the advance requests service facilitates better assignment, routing, and rebalancing, thus improving the effective supply and distribution of vehicles.

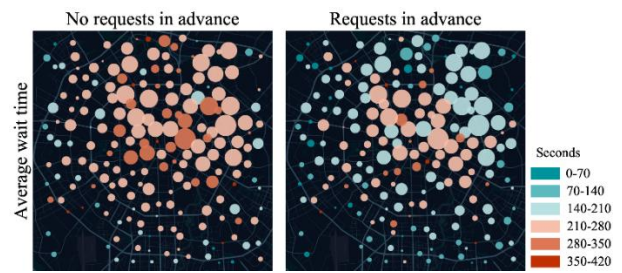

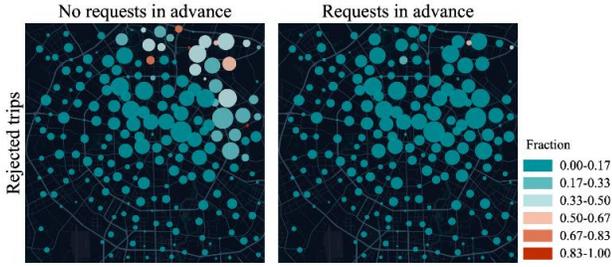

**Fig. 13.** Spatial distribution of wait time (Top) and rejected trips (Bottom) for the case with requests in advance (horizon 30 minutes) and no advance requests. The vehicle capacity is 2.

*Fleet Utilization*. Figure 14 illustrates the required fleet size as a function of LOS constraints and advance requests horizon length. The fleet size used is determined by the number of vehicles that served at least one request (active vehicles). The analysis is for the morning period (7:30-8:30 am), all trips shared, vehicle capacity of 4, and normal traffic conditions).

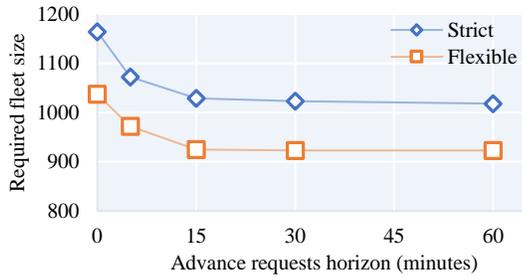

**Fig. 14.** Impact of LOS constraints and advance request horizon on the fleet size required to serve requests. All shared trips, vehicle capacity 4, and normal traffic.

The required fleet size is lower for a system with advanced requests regardless of LOS constraints. Under strict LOS constraints and all requests on-demand ($H = 0$), the number of active vehicles is 1,164. However, with advance requests and a planning horizon $H = 15$, the actual fleet size drops to 1,029, an 11.6% reduction. Using flexible LOS constraints has a significant impact on the required fleet size regardless of the advance requests horizon. With all requests being on demand, the service's fleet size with flexible LOS constraints is reduced by 10.9% compared to the service with strict LOS constraints. With $H = 15$ the reduction is 10.1% (from 1,029 for strict LOS constraints to 925 for flexible). The impact of advance requests horizon on fleet size is diminishing. Actually, the fleet size for $H = 15$ is very close to the fleet size required when the advance requests horizon is 60 minutes. As discussed earlier, with average trip lengths less than 6 miles, an advance requests horizon of 15 minutes captures most of the information useful for matching and scheduling purposes.

Figure 15 shows the distribution of the maximum occupancy of active vehicles. The results show that when the vehicle capacity is relatively small (4 seats or less), most of the trips operate at capacity. However, as the capacity increases, the utilization of the capacity diminishes, with the majority of the trips carrying 4-5 passengers. The results are consistent with the findings in Figure 10, which shows a marginal impact on VMR of increasing capacity beyond 7. Hence, operating a fleet with a capacity larger than that may have diminishing and possibly even negative returns considering higher operation costs, fuel consumption, etc.

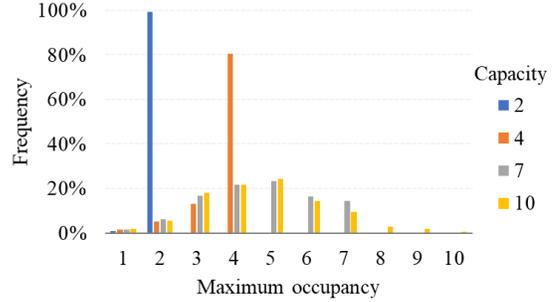

**Fig. 15.** Distribution of the maximum occupancy of active vehicles

## V. Conclusion

The paper introduces an advance requests operating model for the ride pooling problem. Efficient algorithms for scheduling requests and routing/rebalancing vehicles are used to evaluate various system configurations. Using a large-scale TNC dataset, the potential benefits of advance requests are explored under various operating settings (advance requests horizon, vehicle capacity), user preferences (willingness to share and place requests in advance), LOS constraints on maximum waiting, and delay times, and traffic conditions. The key findings include:

- The combined effect of willingness to place requests in advance (even with a short horizon - near-on-demand requests), willingness to share trips, and flexibility in LOS constraints can have an important impact on the efficiency of ride pooling services resulting in a win-win situation for all stakeholders involved (passengers, operators, cities).
- The benefits in terms of VMT savings, fleet size reduction, % of served trips, and lower passenger wait times from such customer flexibility are significant.
- Near-on-Demand services (i.e. a short advance requests horizon of 15 minutes) capture most of the benefits for all involved (except VMR savings that keep improving as the advance requests horizon increases). From a practical point of view, the adoption of such services may be attainable with the right incentives or pricing.
- Increasing willingness to share trips is important in improving sustainability metrics, while advance requests provide additional VMT reduction and improve LOS.
- With small vehicle capacity (equivalent to a service which there is a limit to the number of passengers sharing a trip), advance requests increase the % of trips actually shared.
- Increasing vehicle capacity has diminishing returns when it exceeds a certain value (7 in our case).
- A potentially important finding is that operating with advance requests results in a more equitable distribution of the benefits, with improved service provided to the areas further away from the CBD.

The results provide useful insights into the deployment of ride pooling services and design of pricing strategies to incentivize customers. The concept is applicable to centrally controlled TNC services and autonomous shared MoD systems. The developed ARRP platform is general and can be used to evaluate a wide range of operating strategies under diverse user behavior and uncertainty about future demands.

Pricing plays a critical role in implementing the near-on-demand ride pooling system. Future work may investigate the impact of pricing strategies as a function of the advance requests horizon on user behavior [36]. Other promising research directions include the exploration of other objective functions, investigation of "optimal" model parameters in assignment and rebalancing, and analyzing the trade-offs between users adopting advance requests and reduced VMT.

## VI. Acknowledgments

The authors would like to thank DiDi Chuxing, China for providing the data for this research, the Massachusetts Green High-Performance Computing Center (MGHPCC) for support on computing. We also thank Ms. Yunqing Chen for preparing the graphs in Section IV.

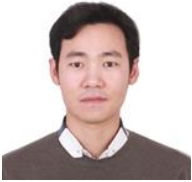
**Zhenliang Ma** is an Assistant Professor at Monash University, Australia. His research interest is at the intersection of optimization, machine learning, and simulation. He focuses on the inference, prediction and design, through the integration of novel data sources into mathematical learning models. The applications include traffic, transit, and future mobility services.

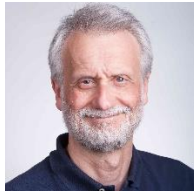
**Haris N. Koutsopoulos** is a Professor with the Department of Civil and Environmental Engineering, Northeastern University, Boston. His current research interests include on the use of data from opportunistic and dedicated sensors to improve planning, operations, monitoring, and control of urban transportation systems. He is the Founder of the iMobility lab, which uses ICT to address urban mobility problems. The Laboratory received the IBM Smarter Planet Award in 2012.